\documentclass[prl,twocolumn,showpacs,preprintnumbers,amsmath,amssymb]{revtex4}

\usepackage{graphicx}
\usepackage{dcolumn}
\usepackage{bm}

\begin{document}

\preprint{PRL/13.06.02}

\title{Substitution induced pinning in MgB$_2$ superconductor doped with SiC nano-particles}

\author{S. X. Dou}
\author{A. V. Pan}
\author{S. Zhou}
\author{M. Ionescu}
\author{H. K. Liu}
\affiliation{Institute for Superconducting and Electronic Materials, University of Wollongong, Wollongong, NSW 2522, Australia}
\author{P. R. Munroe}
\affiliation{Electron Microscopy Unit, University of New South Wales, Sydney, NSW 2052 Australia}

\date{June 13, 2002} 

\begin{abstract}
By doping MgB$_2$ superconductor with SiC nano-particles, we have successfully introduced pinning sites directly into the crystal lattice of MgB$_2$ grains (intra-grain pinning). It became possible due to the combination of counter-balanced Si and C co-substitution for B, leading to a large number of intra-granular dislocations and the dispersed nano-size impurities induced by the substitution. The magnetic field dependence of the critical current density was significantly improved in a wide temperature range, whereas the transition temperature in the sample MgB$_2$(SiC)$_x$ having $x = 0.34$, the highest doping level prepared, dropped only by 2.6~K.
\end{abstract}

 
\maketitle

Since the discovery of superconductivity in MgB$_2$ superconductor \cite{nature1}, many groups have been trying to improve the critical current density ($J_c$) and its magnetic field dependence in various forms of the material. Different sophisticated techniques have been used to enhance pinning, such as oxygen alloying in MgB$_2$ thin films \cite{eom} and proton irradiation of MgB$_2$ powder \cite{bogu}. A simpler method to increase pinning properties of the superconductor is the process of chemical doping. However, despite of numerous attempts results remain controversial as doping was only limited to additions which were ineffective for pinning at temperatures above $T = 20$~K \cite{zhao,feng,wang}, which is considered to be a benchmarking operating temperature for MgB$_2$. In this letter, we present the result of SiC nano-particle doping on critical temperature $T_c$, critical current density, crystal structure, and vortex pinning in MgB$_2$(SiC)$_x$. We clearly demonstrate that despite of a low density ($< 50$\%) and unoptimised composition, $J_c(H)$ and irreversibility field $H_{\rm irr}$ are improved significantly due to Si and C co-substitution for B in the crystal lattice, inducing strong intra-granular pinning in MgB$_2$ grains.

MgB$_2$ pellet samples were prepared by in-situ reaction method which has been previously described in detail \cite{soltan}. Powders of magnesium (99\%) and amorphous boron (99\%) were well mixed with SiC powder with the atomic weight ratio of (Mg+2B)(SiC)$_x$  where $x = 0$, 0.057, 0.115, 0.23 and 0.34 for samples 1 to 5, respectively (Table~\ref{1}). SiC additives consisted of 10-100~nm large particles. The mixed powders were loaded into Fe tubes. The composite tubes were groove-rolled, sealed in a Fe tube and then heated at preset temperatures of 950$^{\circ}$C, for 3 hours in flowing high purity Ar. This was followed by quenching to liquid nitrogen. All measured samples had a rectangular shape having typical dimensions $d \times w \times l = (0.67 \pm 0.15) \times (0.85 \pm 0.15) \times(1.45 \pm 0.10)$~mm$^3$. Table~\ref{1} gives the samples' parameters and selected results on $T_c$, width of the transition $\Delta T$, and $H_{\rm irr}$.

\begin{table}
\caption{\label{1} Selected MgB$_2$(SiC)$_x$ sample parameters.}
\begin{ruledtabular}
\begin{tabular}{cccccccccc}
Sample& SiC &$x$&Density & $T_c$& $\Delta T$& $H_{\rm irr}$& $H_{\rm irr}$& $H_{\rm irr}$ & $H_{\rm irr}$ \\
No & wt\%& &g/cm$^2$& K & K& 10K& 20K& 25~K& 30K\\
\hline
1 & 0 & 0& 1.20& 38.6& 1.25& 5.9& 3.9& 3.6& 2.2\\
2 & 5 & 0.057& 1.21& 37.8& 1.29& 7.4& 5.1& 4.5& 2.5\\
3 & 10 & 0.115& 1.22& 37.4& 1.75& 7.6& 5.8& $>5.0$& 2.5\\
4 & 20 & 0.23& 1.17& 36.9& 2.28& 5.6& 4.4& 3.7& 1.5\\
5 & 30 & 0.34& 1.30& 36.0& 2.48& 5.3& 4.0& 3.5& 1.9\\
\end{tabular}
\end{ruledtabular}
\end{table}

The magnetization as a function of temperature $T$ and magnetic field $H$ applied along the longest sample dimension was measured using a Quantum Design Magnetic Property and Physical Property Measurement Systems within the field range $|H| \le 9$~T, and within the temperature range of $5~{\rm K} \le T \le 30$~K.  A magnetic $J_c$ was derived from the half-width of magnetization loops $\Delta M$, using the following critical state formula: $J_c = k10 \Delta M/d$, where $k = 12w/(3w-d)$ is a geometrical factor.

\begin{figure}
\includegraphics[scale=0.28]{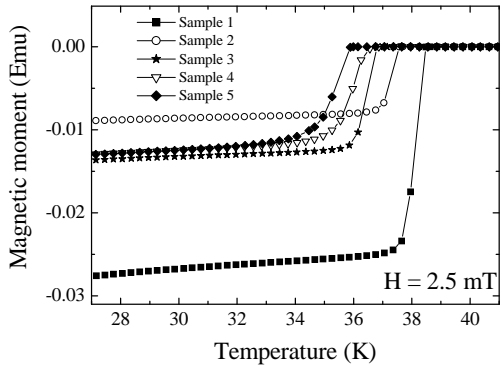}
\caption{\label{tc}Magnetization measurements as a function of temperature for all samples. $T_c$ decreased only by 2.6~K for the highest doping level sample~5.}
\end{figure}

Fig.~\ref{tc} shows the magnetic moment behavior as a function of temperature. The transition temperature $T_c$ onset for the undoped sample (38.6~K) is similar to the values reported by other groups. For the doped samples, $T_c$ decreases (Fig.~\ref{tc}) whereas transition width $\Delta T_c$ increases (Table~\ref{1}) with increasing doping level. It is striking to note that despite the large amount of non-superconducting phases added, the $T_c$ droped only by 2.6~K for the highest SiC doping level (Sample~5). In contrast, $T_c$ was depressed by almost 10~K for 7at\% C substitution of B in MgB$_2$ \cite{takennobu} and by about 0.5~K for 0.5at\% Si substitution \cite{cimberle}. These results suggest that the higher $T_c$ tolerance of MgB$_2$ to SiC doping is attributable to the co-doping of C and Si because the average size of C (0.091~nm) and Si (0.146~nm) is almost perfectly matches that of B (0.117~nm). It is evident that the co-doping with SiC counter-balanced the negative effect on $T_c$ of the single element doping. As a result the superconducting energy gap throughout the superconductor decreases only a little even at the highest doping level introduced ($x = 0.34$). However we anticipate some very local gap variation due to the substitution. In order to determine precisely the maximum level of SiC substitution for B in MgB$_2$, a study of MgB$_{2-x}$(SiC)$_{x/2}$ and MgB$_{2-x-y}$Si$_x$C$_y$ is underway and results will be reported  elsewhere.

\begin{figure}
\vspace{-0.5cm}
\includegraphics[scale=0.45]{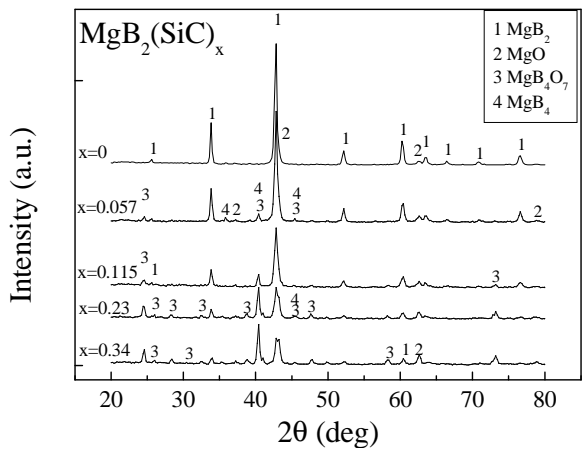}
\vspace{-6.5cm}
\caption{\label{xrd}XRD patterns for the undoped and SiC-doped samples. Note that the MgB$_4$, MgO and MgB$_4$O$_7$ peaks increased significantly with the increasing amount of SiC.}
\end{figure}

\begin{figure}
\includegraphics[scale=0.28]{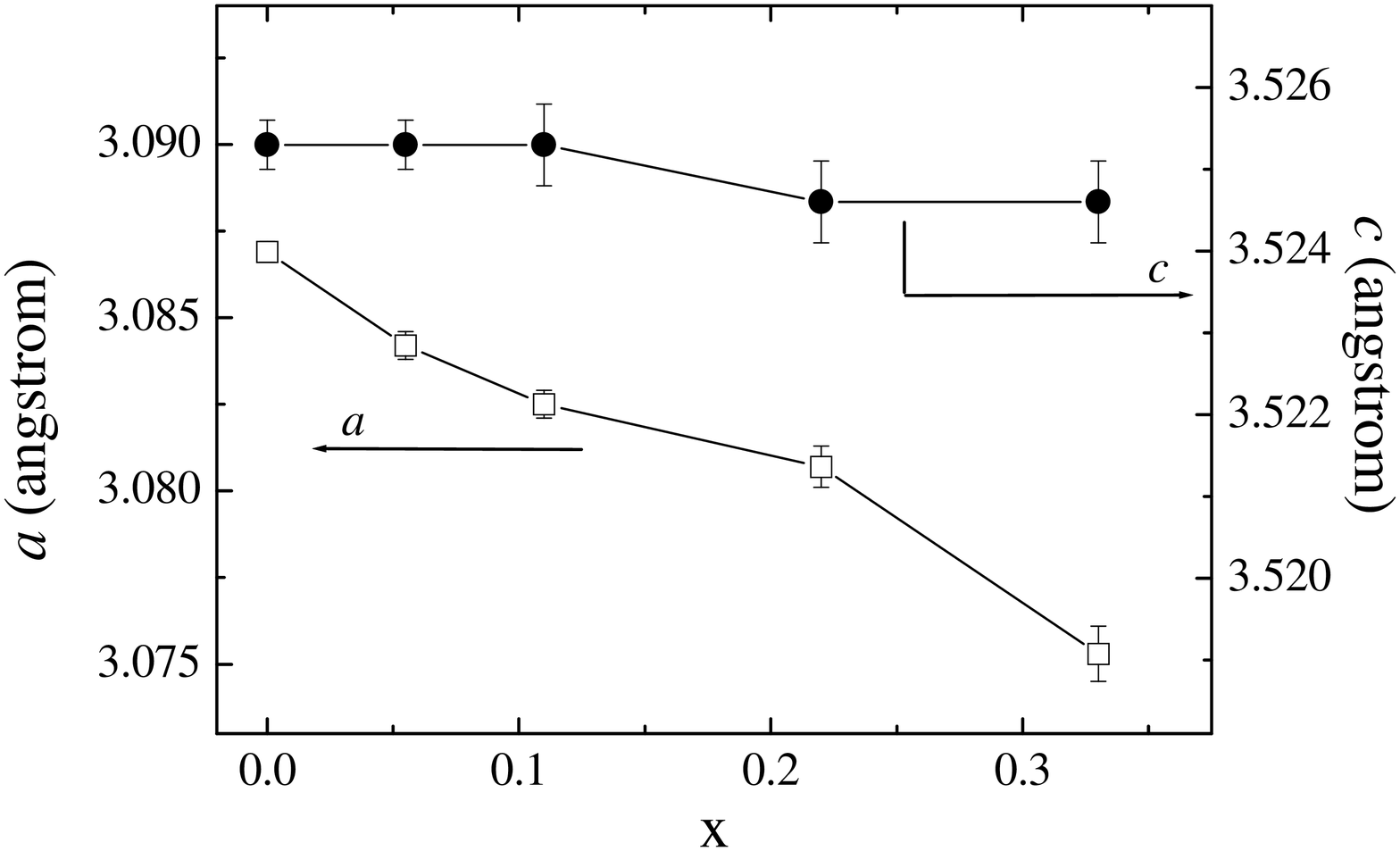}
\caption{\label{cl}Lattice parameters $a$ and $c$ as a function of the SiC content $x$. Note $a$ is decreasing with increasing SiC dopant, and does not reach saturation at $x = 0.34$.}
\end{figure}

Fig.~\ref{xrd} shows X-ray diffraction (XRD) patterns for the SiC doped and undoped samples. The X-ray scans were recorded using Cu$_{K\alpha} = 1.5418$~angstrom, and indexed within the space group P6/mmm. For the in-phase reflections which occur in Fig.~\ref{xrd} between $2\theta = 33^{\circ}$ and $2\theta = 34^{\circ}$ indexed as (100), the centroid of the peak clearly shifts to higher $2\theta$ values with increasing $x$. In the same time the centroid of the peaks which occur between $2\theta = 51^{\circ}$ and $2\theta = 52^{\circ}$, and indexed as (002), the shift to higher $2\theta$ values with increasing $x$ is marginal within the experimental error. The lattice parameters, $a$ and $c$ of the hexagonal AlB$_2$-type structure of MgB$_2$ were calculated using these peak shifts as shown in Fig.~\ref{cl}. The continuous decrease of $a$ with increasing SiC doping level indicates that B is substituted by C and Si. The total change of $a$ from $x = 0$ to $x = 0.34$ is 0.012~angstrom, comparing with single element doping such as C \cite{takennobu}, where $a$ reached a plateau at C content of 7at\% of B, the decrease of $a$ was 0.016~angstrom. This indicates that co-doping of Si and C into MgB$_2$ substantially reduced the change of $a$ and raised the saturation level due to the counter-balance effect of Si and C. This provides an additional support to our explanation of the small $T_c$ degradation with increasing amount of the SiC dopant. Moreover, the amount of the non-superconducting phases increases with increasing SiC doping level, and at $x \ge 0.23$ it even exceeds the amount of MgB$_2$ (Fig.~\ref{xrd}). The appearance of the non-superconducting phases, MgO, MgB$_4$, MgB$_4$O$_7$, can also be attributed to the substitution of B positions by Si and C, resulting in excess of B. Some extra B was incorporated into MgO to form MgB$_4$O$_7$. The extra oxygen may be brought in by the SiC dopant which absorbed moisture or oxygen during storage. Note, there are no SiC peaks indexed even up to $x=0.34$.

\begin{figure}
\includegraphics[scale=0.25]{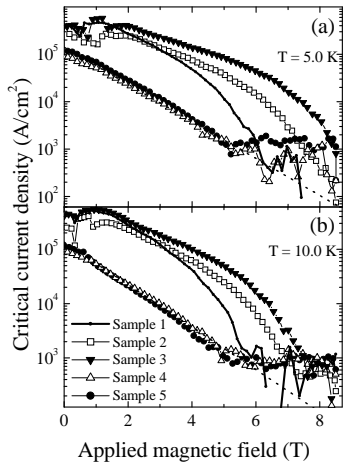}
\caption{\label{jc}Critical current density as a function of magnetic field at 5~K (a) and 10~K (b) for all samples.}
\end{figure}

Fig.~\ref{jc} shows the $J_c(H)$ curves for doped and undoped samples at 5~K (a) and 10~K (b). These results show the following striking features. The $J_c(H)$ curves for undoped sample~1 show a crossover with those for doped samples~2 and 3 at higher fields. $H_{\rm irr}$ for doped sample~3 is 5.8~T at 20~K and 7.6~T at 10~K, compared to 3.9~T and 5.9~T for the undoped one at the same temperatures. Although SiC doping at $x > 0.115$ caused a reduction of $J_c$, it is important to note that the $J_c$ for all the doped samples drops with increasing field much slower than for the undoped one, in particular, at higher fields. Furthermore, $J_c$ curves for the doped samples~4 and 5 show an exponential behavior (dotted lines) as a function of the magnetic field up to the measurement noise level at high fields, while the $J_c$ curve for the undoped sample shows a rapid downward bending. At $T = 5$~K and 4~T, $J_c$ for the doped sample~3 reaches $1.4\times10^5$~A/cm$^2$, and at 6~T it is $\sim 30$ times larger than $J_c$ of the undoped sample~1. It is also interesting to note that at lower temperatures ($T < 10$~K) $J_c(H)$ for higher doping level samples (4 and 5) declines faster than those for low doping samples (2 and 3). However, at higher temperatures ($T \ge 10$~K) samples with higher doping level exhibit almost parallel $J_c(H)$ curves, while at 30~K, sample~5 showed even slower $J_c$ drop in increasing field than all others. This $J_c(H,T)$ behavior indicates a stronger pinning enhancement effect at high temperatures than at low temperatures. It should be added that in order to reduce flux-jumps effect \cite{dou} at low fields (Fig.~\ref{jc}) we measured relatively small samples which had a drawback that the measurements were less sensitive in vicinity of $H_{\rm irr}$.
 
\begin{figure}
\includegraphics[scale=0.34]{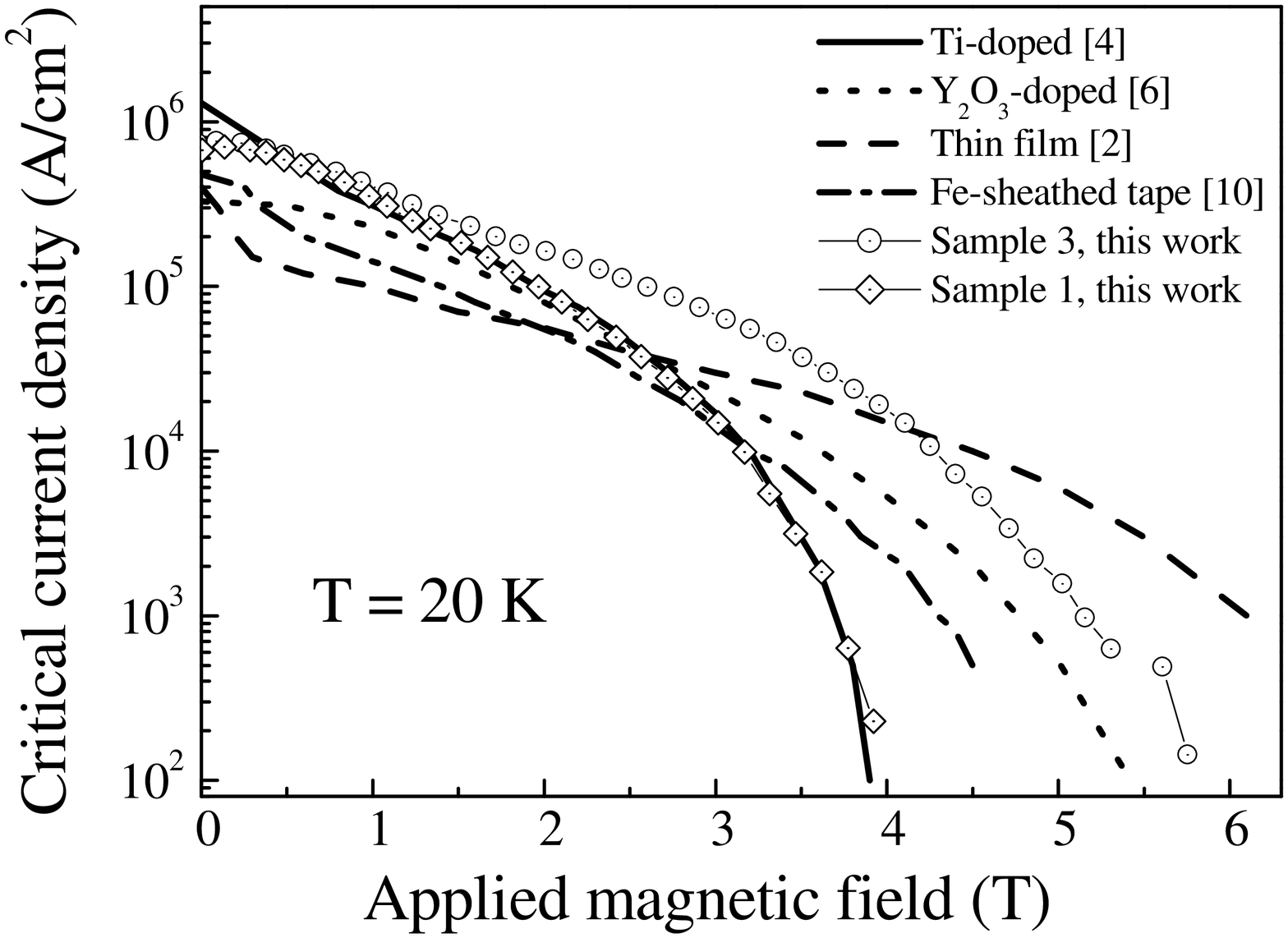}
\caption{\label{jccomp} A comparison of $J_c(H)$ for the reference undoped sample~1 and SiC doped sample~3 (this work) with those for Ti \cite{zhao} and Y$_2$O$_3$ doped \cite{wang} bulk samples, thin film with strong pinning \cite{eom} and Fe-sheathed tape \cite{beneduce}, representing the state-of-the-art performance of MgB$_2$ in various forms.}
\end{figure}

Fig.~\ref{jccomp} shows a comparison of $J_c(H)$ behavior for undoped sample~1 and doped sample 3 with data reported in literature at 20~K. It is evident that despite the low density (Table~\ref{1}), which is $< 50$\% of the theoretical density, and unoptimised composition at this stage, the $J_c$ for the SiC doped sample drops slower than $J_c$ for other element doped samples \cite{zhao,feng,wang}, the best Fe/MgB$_2$ tape \cite{beneduce}, and is close to $J_c(H)$ behavior of the thin film with strongest pinning reported so far \cite{eom}. At 20~K,  sample~3 has $J_c$ value of 18,000~A/cm$^2$ at 4~T, which is two order of magnitude higher than the undoped reference sample~1. This is higher by a factor of 8 than that of the state-of-the-art Fe/MgB$_2$ tape \cite{beneduce}, and comparable to Ta doped MgB2 made using high pressure synthesis (2GPa) \cite{prikhna}. These are the best $J_c$ values ever reported for bulk and wires made under normal conditions. It should be pointed out that since the density of our samples is low, making the super-current paths highly percolative, $J_c$ values are far from optimum. We can anticipate that higher $J_c$ can be achieved if the density of the samples is further improved.

Regarding the mechanism of the enhancement of $J_c$ at higher fields, it is necessary to recognize the special features of SiC doping. First, in contrast to previous work on doping for improving $J_c$ \cite{zhao,feng,wang,prikhna}, SiC doping has no densification effect, independent of doping level. This is understandable because SiC has very high melting point and would not act as sintering aid at the temperature range of 800$^{\circ}$C to 950$^{\circ}$C. Second, SiC doping takes place in the form of the substitution while in the previous work the element doping is in the form of additives \cite{zhao,feng,wang,mickelson,takennobu,cimberle,prikhna}, either not incorporated into the crystal lattice or incorporated to a lesser extent than in our work. Doping MgB$_2$ with Ti and Zr showed an improvement of $J_c$ in self field and 4~K \cite{zhao,feng}. However, there is no evidence for improved pinning as $J_c$ drops rapidly with increasing field ($H_{\rm irr} = 4$~T at 20~K). Doping MgB$_2$ using Y$_2$O$_3$ nanoparticles showed an improved irreversibility field at 4.2~K, but $H_{\rm irr}$ for the doped samples is not as good as the undoped ones at 20~K \cite{wang}. Further, separate doping with a small amount of Li, Al and Si showed some increase in $J_c$, but there was no improvement in $H_{\rm irr}$ \cite{cimberle}. It is evident that the additive pinning is more effective at low temperatures mainly due to densification effect, while the additives at the grain boundaries decouple the grains at high temperatures.

\begin{figure}[t]
\includegraphics[scale=0.37]{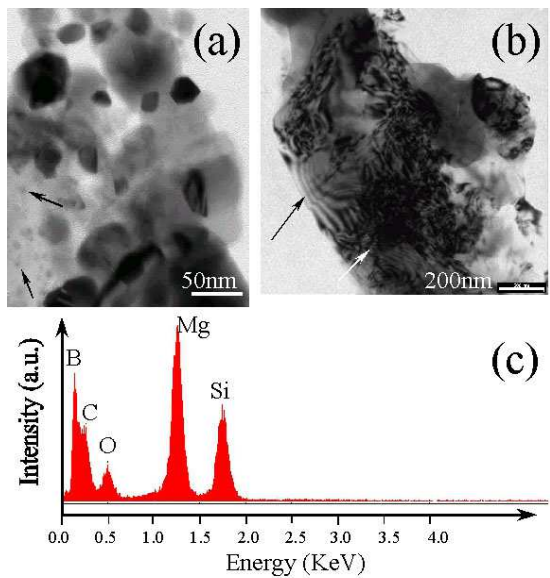}
\caption{\label{tem} TEM images (a,b) and EDX analysis (c) of the doped sample 3 (a,b). 20-500~nm large MgB$_2$ grains are shown in both TEM images. Impurities particles are $< 10$~nm (shown by arrows in (a)). A large number of dislocations (shown by arrows in (b)) is present in the MgB$_2$ grains. The EDX analysis for one of these grains shows that both Si and C are present within the grain (c).}
\end{figure}

Thus, in the SiC doped samples we suggest strong induced intra-granular pinning mechanism which is particular important due to the fact that the inter-granular weak-link effect is negligible in MgB$_2$ \cite{nature2}. The strong pinning can be attributed to be due to B substitution by Si and C, leading to (i) local superconducting order parameter fluctuations and to (ii) crystal lattice defect (dislocations) formation, both would act as effective (intra-granular) pinning sites intrinsic in nature. A considerable contribution to overall pinning might also be introduced by impurities formed due to the substitution. Indeed, the high content of MgO and other impurity phases in the SiC doped samples could also be potential pinning centers, consistent with the results obtained for the thin film with strong pinning where the ratio of Mg:B:O reached 1.0:0.9:0.7 \cite{eom}. When SiC reacts with liquid Mg and amorphous B at the sintering temperatures, the nanoparticles may act as nucleation sites to form MgB$_2$ and other phases. Some nanoparticles can be included within the grains as inclusions. Thus, the reaction induced products are highly dispersed in the bulk matrix. These arguments are strongly supported by study of microstructures with a help of transmission electron microscopy (TEM) and energy dispersive X-ray spectroscopy (EDX) analysis (Fig.~\ref{tem}). The EDX analysis results exhibited that the Mg:Si ratio is identical across the entire sample area, pointing out a complete level of the substitution and a homogeneous phase distribution (Fig.~\ref{tem}(c)). SiC particles are absent in TEM images, again indicating the fact that they are completely dissolved in MgB$_2$.  Local distribution analysis of the elements within grains of 20-500~nm large (Fig.~\ref{tem}(a,b)) showed that Mg and B appeared to be uniformly distributed, however there are local variations in Si, O and C which might suggest very local differences in substitution. Further, the SiC-"solute" cause local lattice strains in MgB2-"solvent" creating not only local fluctuation of the superconducting order parameter, but also a number of dislocation clearly seen in Fig.~\ref{tem}(b). Both, the fluctuations and the dislocations, strongly contribute to overall pinning. Some smaller particles $< 10$~nm found in Fig.~\ref{tem}(a) are impurities of MgO and borides identified by XRD pattern (Fig.~\ref{xrd}). Thus, the results of the present work suggest that a combination of the presumably dominating substitutional effect, possibly aided by the highly-localized non-uniform nature of these substitutions, and highly dispersed additives induced through the substitution is responsible for the enhanced flux pinning in SiC-doped MgB$_2$.

In summary, we have demonstrated B substitution effect in MgB$_2$ caused by SiC nano-particles doping. Due to the counter-balance effect of Si and C atom sizes, in contrast to single element doping, the doping has the following influence on properties of MgB$_2$(SiC)$_x$: (i) only a small reduction of the crystal lattice parameters up to the highest investigated level of doping ($x = 0.34$). (ii) $T_c$ was reduced by only $\le 2.6$~K for the entire doping range. (iii) The high level of co-substitution induced defects (dislocations) and local composition change within superconducting grains act as effective intra-granular pinning centers. (iv) High $J_c$ and improved $H_{\rm irr}$ were achieved within wide field and temperature ranges. There is ample room for further improvement if density of the composition is increased. We anticipate that for practical applications MgB$_2$ superconductors will be made using the formula MgB$_x$Si$_y$C$_z$ where $x + y +z \ge 2$, instead of pure MgB$_2$. 

\begin{acknowledgments}
We thank E. W. Collings, J. Horvat, T. Silver, M.J. Qin, M. Sumption, M. Tomsic, X.L. Wang for their helpful discussions. This work was supported by the Australian Research Council, Hyper Tech Research Inc OH USA and the University of Wollongong.
\end{acknowledgments}

\end{document}